\documentstyle[a4,11pt,epsf,twoside]{article}
\newcommand{\gsim}{\mbox{ \raisebox{-1.0ex}{$\stackrel{\textstyle >}
{\textstyle \sim}$ }}}
\newcommand{\lsim}{\mbox{ \raisebox{-1.0ex}{$\stackrel{\textstyle <}
{\textstyle \sim}$ }}}

\begin{document}
\title{THEORETICAL INTRODUCTION TO LINEAR COLLIDER PHYSICS
       \thanks{Talk given at International Workshop on Linear Colliders,
August 26 - 30, 2002, Jeju Island, Korea. This work  was supported in part by 
Grand-in-Aid for Scientific Research, (C) (No. 13640309). 
	}
}
\author{YASUHIRO OKADA$^{1,2} $
	\thanks{e-mail address: yasuhiro.okada@kek.jp}
\\
\\
        $^1$ {\it Institute of Particle and Nuclear Studies, KEK, Japan}
\\
        $^2$ {\it Department of Particle and Nuclear Physics,}
\\
 {\it Graduate University of Advanced Studies, Japan} 
}
\date{}
\maketitle
\vspace{-10cm}
\begin{flushright}
KEK-TH-884    \\
April 2003  \\
\end{flushright}
\vspace{7.5cm}
\vspace{-0.7cm}
\begin{abstract}
Physics at an $e^+e^-$ linear collider is described as
a theoretical introduction to Linear Collider Workshop 2002.

\end{abstract}

\section{Introduction}
There have been remarkable progresses in particle physics in the last
century. New discoveries have changed our view on elementary particles
and cosmology. In our current standpoint, goals of the elementary particle 
physics are not restricted to finding fundamental constitutes of matter and 
fundamental interactions. We have to address questions concerning
the vacuum state of the Universe and the structure 
of space and time to understand the law of elementary particles.
These issues are also closely related to questions like 
how the Universe have begun and evolved to today.

Current understanding of particle physics is based on the Standard Model (SM),
which is a gauge theory of quarks and leptons. The gauge groups
are  SU(3) $\times$ SU(2) $\times$ U(1): The SU(3) group describes
the strong interaction, and the SU(2) and U(1) groups correspond to
the electromagnetic and the weak interactions. The SM has 
been tested experimentally for the last three decades, and especially the 
gauge interaction among fermions and bosons has been studied very precisely
at LEP and SLC experiments in 1990's. The gauge structure of the SM
is now well established experimentally. 

The next important steps of the particle physics are:
\begin{itemize}
\item To establish the mechanism of the electroweak symmetry breaking 
and the particles' mass generation .
\item To find the physics scenario beyond the electroweak scale.
\end{itemize}
 
I would like to discuss how $e^+e^-$ linear collider (LC) experiments 
can play essential roles for these purposes. Studies of research potentials 
at an $e^+e^-$ LC have been carried out in world-wide, and  
its outcome is found in literatures, especially in recent 
reports~\cite{Abe:2001wn, Aguilar-Saavedra:2001rg, Abe:2001gc}.

\section{Higgs physics}
Higgs fields are introduced to break 
electroweak symmetry  and generate masses for elementary particles.
Although only one Higgs doublet field is necessary for this 
purpose, little is known about the structure of the Higgs sector.
We do not know how many Higgs fields exists, whether there are
fields other than weak doublet fields, and what is the mass of
the Higgs particle. We have little information
on couplings between the Higgs field and fermions/gauge bosons.
We need the LC experiment to answer these questions.

\subsection{Higgs boson mass}
The Higgs boson mass is the most important parameter in the Higgs sector.
Since particle mass is generated from interaction with the Higgs 
field in the SM, the mass of the particle carries information 
on the strength of its interaction to the Higgs field. This is also true for 
the Higgs boson mass itself. In particular, the Higgs-boson mass in the 
SM is expressed as $m_h = \sqrt{2\lambda} v$, where 
$\lambda$ is the Higgs self-coupling constant 
($V = -\mu^2|\phi|^2+ \lambda |\phi|^4$).
In general, a light Higgs boson corresponds to a weakly-coupled
scenario for the mechanism of the electroweak symmetry 
breaking. Supersymmetry (SUSY), grand unified theory, and string unification
are such examples. On the other hand, a heavy Higgs boson is a signal
that some strong dynamics is behind the symmetry breaking. 

The current experimental lower bond on the Higgs boson mass in the Standard
Model is 114 GeV from LEP experiments at 95 \% CL. On the other hand,
from global analysis of the SM, we can derive the 95 \% CL 
upper bound as 193 GeV~\cite{Grunewald:ICHEP2002}. 
Although this bound is valid for the SM Higgs boson, a 
weakly-coupled scenario is favored for the 
dynamics of the electroweak symmetry breaking.

If we introduce some theoretical assumption, a possible mass 
range of the Higgs boson can be derived. For instance, if the
SM is assumed to be valid without any change,
except for a possible see-saw mechanism for neutrino mass generation,
the upper and lower bounds of the Higgs boson mass are determined 
as a function of the cut-off scale $(\Lambda)$ of the theory.   
For  $\Lambda= 10^{19}$ GeV of the Planck scale, the lower and upper 
bounds are about 130 GeV and 180 GeV, respectively. The Higgs sector of the 
minimal supersymmetric standard Model (MSSM) consists of two Higgs doublets.
A remarkable feature of this model is that the upper bound of the lightest 
CP-even Higgs boson can be derived without reference to the cutoff scale.
For reasonable choice of parameters in the model, especially the 
stop mass and mixing parameters, the bound is given 
by 130 GeV~\cite{Okada:1990vk}.
For other models like general two Higgs doublets and SUSY
models with extended Higgs sectors, we can derive the mass range 
of a Higgs boson whose properties are similar to the SM Higgs boson,
if the cutoff scale is specified~\cite{Kanemura:1999xf, Espinosa:1998re}. 
The mass range is shown in Figure 1
for various models in the case of $\Lambda= 10^{19}$ GeV.
We can see that at least one Higgs boson must exist near or less that 200 GeV
under this theoretical assumption.

\begin{figure}[ht]
\begin{center}
\leavevmode
\epsfxsize=8cm
\epsfysize=6cm
\epsffile{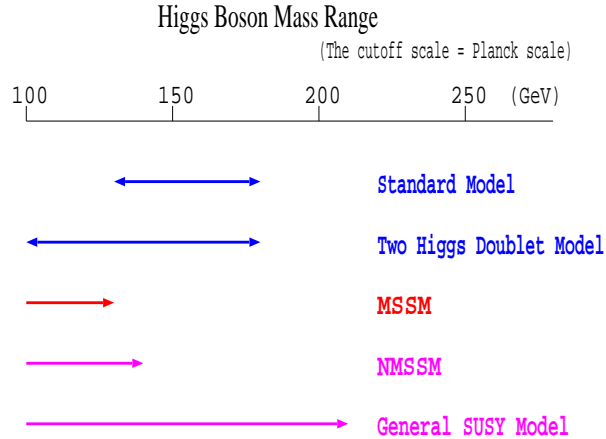}
\caption{Mass range of a Higgs boson for various models. The upper and lower
bound of the Higgs boson mass is derived from an assumption that each model
is valid up to the cut-off scale of the theory, which is taken to be 
$10^{19}$ GeV. For the MSSM case, the mass bound is obtained without
reference to the cutoff scale.
 }
\end{center}
\end{figure}

The Higgs boson search is currently undertaken at Fermilab TEVATRON.
With an integrated luminosity of 15 fb$^{-1}$, it will become evident
whether or not a Higgs boson exists for a relatively small 
mass range like 120 GeV. One of main goals of the LHC experiment, which
is scheduled to start in 2007, is the discovery of a  Higgs boson,
irrespectively of its mass. The Higgs boson discovery is relatively easy 
if its mass exceeds  200 GeV, because the $h\rightarrow ZZ \rightarrow llll$ 
mode is available. For a smaller Higgs boson mass, the Higgs boson search 
involves other modes such as two photon and $W^{(*)}W^{(*)}$ decay modes, and 
$t\bar{t} h$ production. It is believed that the discovery of 
a SM Higgs boson is possible at LHC experiments with a low
integrated luminosity of an order of 10 fb$^{-1}$. In addition, recent studies 
of prospects of the Higgs physics at LHC show that useful information 
on Higgs coupling constants will be obtained in the era of a high 
luminosity run $( \gsim 100 \rm{fb}^{-1})$, especially for a relatively
light Higgs boson. Higgs boson production processes through 
a vector boson fusion play an essential role for this 
purpose~\cite{Zeppenfeld:2002ng}.

\subsection{Higgs physics at linear collides}
Goals of Higgs physics at LCs are to establish 
the mass-generation mechanism for elementary particles and
to clarify the dynamics of the electroweak symmetry breaking.
With an integrated luminosity of 500 fb$^{-1}$, over 100,000
light Higgs bosons can be produced at a LC, and
precise information on various couplings related to the Higgs boson
will be obtained from measurements of production cross section and
decay branching ratios. The LC is a Higgs factory
in this sense.

{\bf Production of Higgs bosons}

For a LC with center-of-mass energy of up to 500 GeV, 
we can detect the SM Higgs boson up to 400 GeV. 
The discovery of a Higgs boson only requires an integrated luminosity 
of a few fb$^{-1}$, if the Higgs boson mass is less than 200 GeV.
If we consider models beyond the SM, 
the production cross section and decay modes may be different from
the SM. For example, the production cross section of the lightest 
CP-even Higgs boson can be reduced for the MSSM and the next-to-minimal
supersymmetric standard model (NMSSM), where an additional gauge-singlet
Higgs field is introduced to the MSSM. However we can show quantitatively
that at least one of Higgs bosons has a production cross section 
in the $e^+e^- \rightarrow Z h_i$ precess which is smaller 
only by a factor of two or three than that of the SM Higgs boson.
Therefore, discovery of a Higgs boson is guaranteed in these 
models~\cite{Kamoshita:1994iv}.

{\bf Spin and parity of the Higgs boson}

In order to confirm that the observed particle has right properties
for the Higgs boson, we first have to determine its spin and
parity. The LC experiment 
is necessary for unambiguous determination of these basic quantum numbers.
We can distinguish the CP property of a spin-0 particle from
the angular distribution of the production angle. Furthermore,
an energy scan at the production threshold region provides
enough information to discriminate the spin and parity of the observed 
particles~\cite{Miller:2001bi}. An integrated luminosity of 
several ten's fb$^{-1}$
are necessary for these determinations.

{\bf Mass-generation mechanism}

In the SM, all quarks, leptons and gauge bosons are massless,
if there is no electroweak symmetry breaking. Masses of these particles are 
generated through interactions with the Higgs field, which is supposed to 
condense in the vacuum. Since the Higgs boson is a physical fluctuation 
mode around the vacuum expectation value, we need to determine the 
Higgs-fermion, Higgs-W-W, Higgs-Z-Z coupling constants, and compare them
with values predicted from the masses of elementary particles, in order
to test the mass generation mechanism.
From the production cross sections of $e^+e^- \rightarrow Zh$ and
$e^+e^- \rightarrow \nu \bar{\nu} h$ processes and Higgs decay 
branching ratios, these coupling constants can be determined precisely.
For a Higgs boson mass of less than 140 GeV, the  $hWW$, $hZZ$, 
$hbb$, $h\tau\tau$ coupling constants can be determined at accuracy of
a few \% level with an integrated luminosity of 500 fb$^{-1}$ at
$\sqrt{s}= 300 - 500$ GeV~\cite
{Abe:2001wn, Aguilar-Saavedra:2001rg, Abe:2001gc}. 
We can also derive the total width of 
the Higgs boson with an error of 5 - 10 \%.  The top Yukawa coupling 
constant can be determined from  $e^+e^- \rightarrow t \bar{t}h$ 
production process, and the center-of-mass energy larger than 700 GeV
is preferable for this purpose. In Figure 2, the mass and coupling 
constant relation is plotted for various particles. In general,
the mass and coupling constant have a relation such as $m_i=\kappa_i v$
in the SM, where $v$ is the vacuum expectation value
of the Higgs field, and the $\kappa_i$ is a dimensionless coupling
constant which can be obtained from a three-point coupling measurement. 
In this figure, the expected statistical
error of each coupling measurement at LC experiments
is shown for charm, $\tau$, bottom, W, Z, and top quark. For the Higgs 
boson, the error is derived from the measurement of the triple Higgs-boson
coupling constant.
\begin{figure}[ht]
\begin{center}
\leavevmode
\epsfxsize=8cm
\epsfysize=8cm
\epsfbox{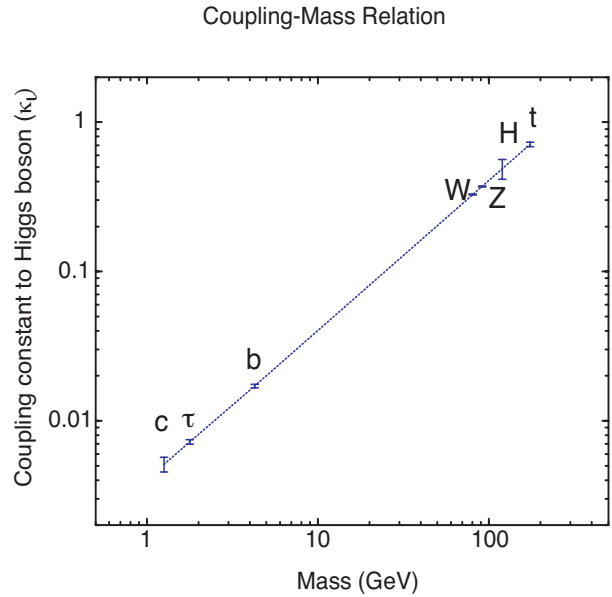}
\caption{Test of mass-generation mechanism performed at LC
experiments. Estimated statistical errors on $\kappa_i$ from
Higgs boson couplings to various particles are shown, where
$m_i=\kappa_i v$ holds in the SM. The Higgs boson mass
is assumed to be 120 GeV. An integrated luminosity is 500 fb$^{-1}$
and the center-of mass energy is $\sqrt{s} = 300$ GeV 
except for the Higgs self-coupling ($\sqrt{s} = 500$ GeV) and 
the top Yukawa coupling ($\sqrt{s} = 700$ GeV) measurements.
 }
\end{center}
\end{figure}

{\bf Higgs self-coupling constant}
 
The first information about the Higgs potential will be obtained 
from the measurement of the triple Higgs-boson coupling constant. 
For this purpose, we need to study double Higgs production processes,
namely, $e^+e^- \rightarrow Zhh$ and 
$e^+e^- \rightarrow \nu \bar{\nu}hh$. In both of these processes, 
in addition to the diagram involving the triple Higgs-boson vertex,
there are diagrams which only depend on vertexes with gauge bosons
and Higgs boson(s). The production cross section is not simply 
proportional to the square of the self-coupling constant.
For $\sqrt{s}= 500$ GeV, the $e^+e^- \rightarrow Zhh$ has much
larger cross section than the $e^+e^- \rightarrow \nu \bar{\nu}hh$ process.
For a higher energy, the production cross section of the
latter process increases, which is an important future of the
WW fusion process, whereas that for the former process decreases as
$1/s$. For $\sqrt{s}\gsim$ 1 TeV, the WW fusion process become more
important for determination of the triple Higgs-boson coupling constant.
It is expected that the sensitivity of the coupling measurement
is about 20 (7) \% for 1 (5) ab$^{-1}$ at $\sqrt{s}=$ 500 (3000) 
GeV~\cite{Battaglia:2001nn}.
The center-of mass energy dependence of the estimated sensitivity 
on the coupling constant is shown in Figure 3, assuming 100 \% 
efficiency for 1 ab$^{-1}$~\cite{Yasui:2002se}.
For $\sqrt{s}\gsim$ 1 TeV, the sensitivity is further improved by using
initial electron (and positron) polarization, because the production cross
section of the WW fusion process is increased by a factor of 2 (4) with
the electron (electron and positron) polarization.  
\begin{figure}[ht]
\begin{center}
\raisebox{6cm}{(a)}
\epsfxsize=5cm
\epsfysize=4cm
\epsffile{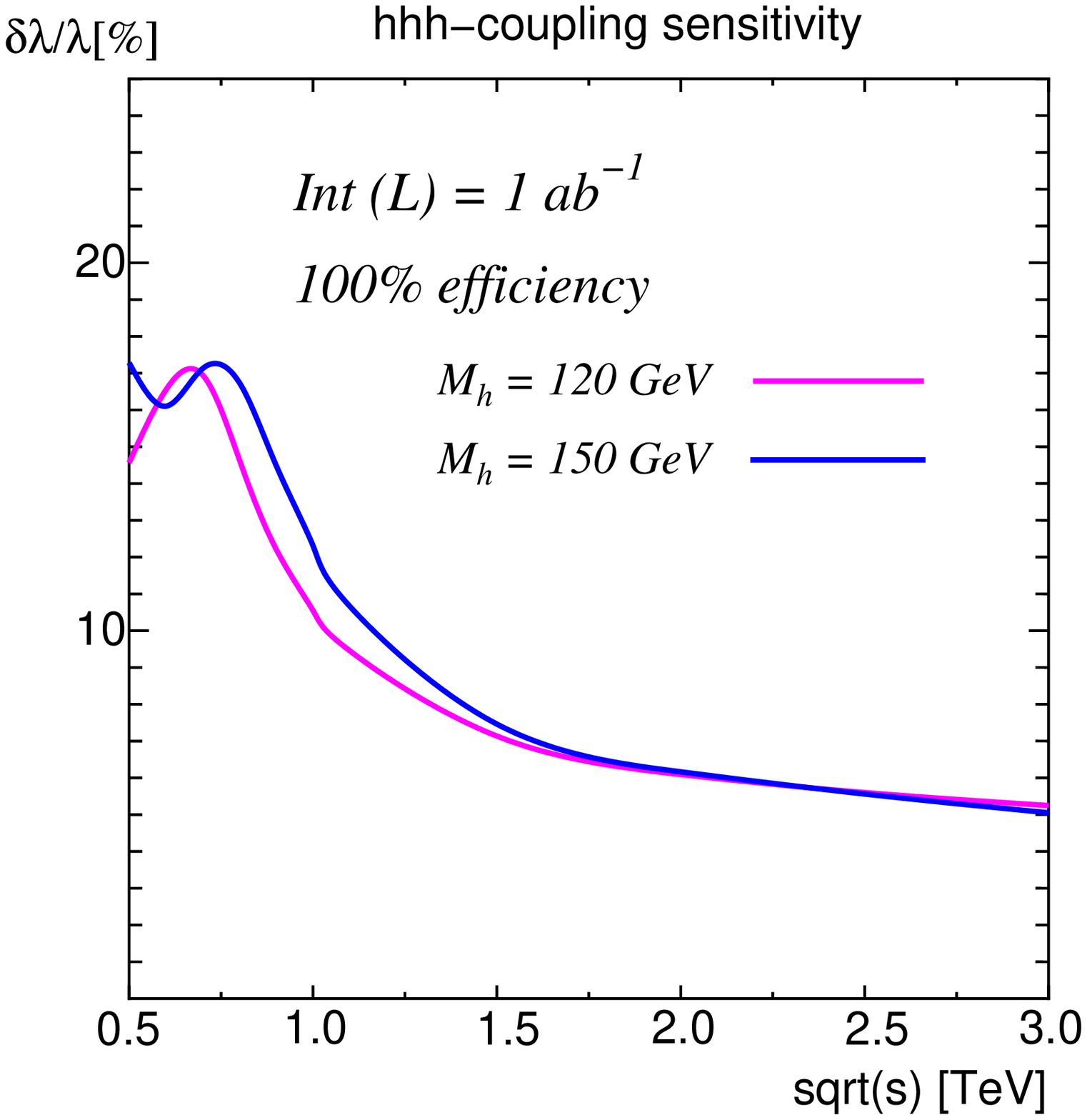}
~~~~~
\raisebox{6cm}{(b)}
\epsfxsize=5cm
\epsfysize=4cm
\epsffile{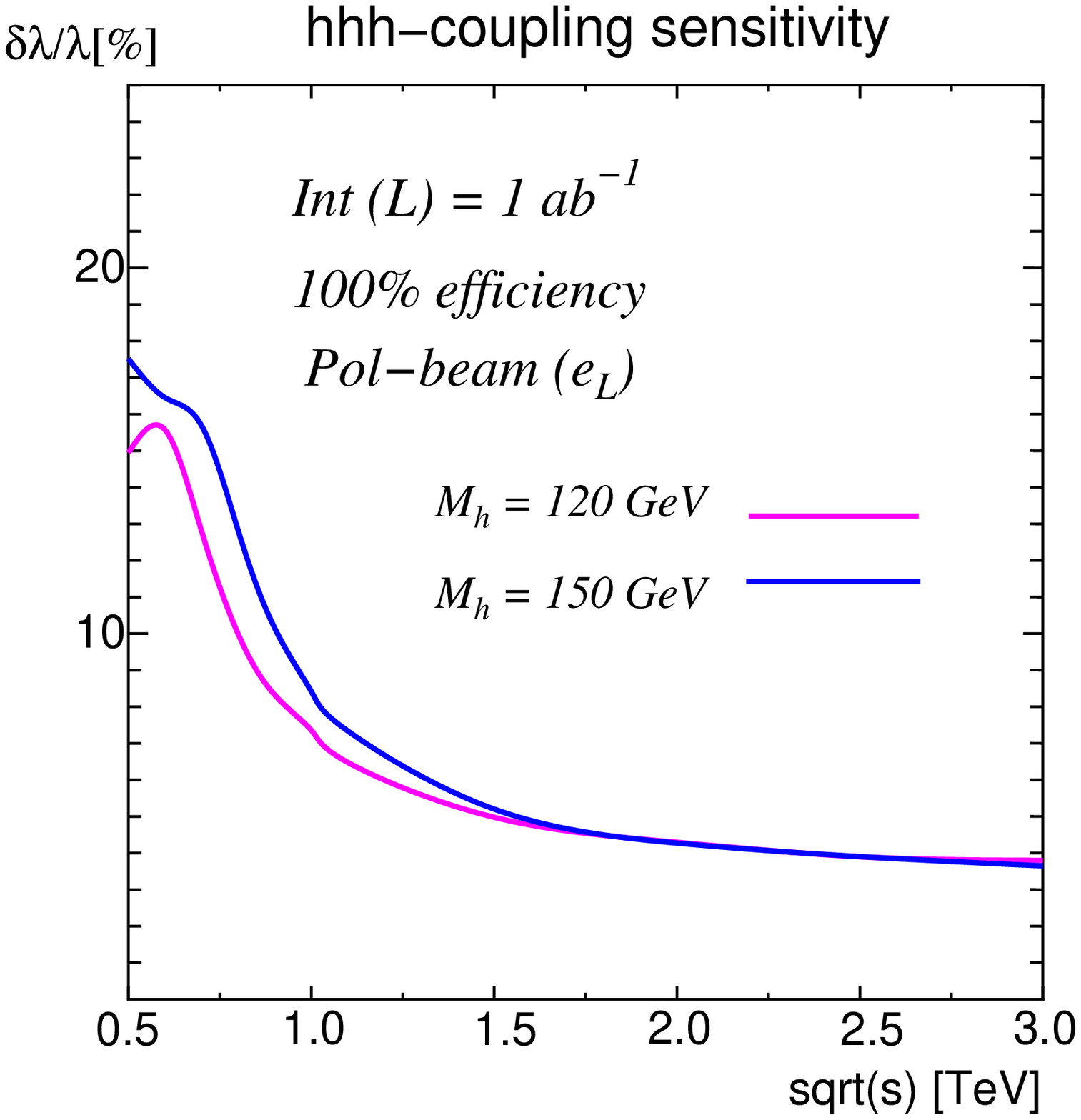}
\caption{
Center-of-mass energy dependence of the sensitivity 
of the triple Higgs-boson coupling constant at LCs.
The integrated luminosity is 1 ab$^{-1}$ without efficiency corrections,
without electron polarization (a) and with 100 \% electron polarization (b)
~\cite{Yasui:2002se}.
 }
\end{center}
\end{figure}

\subsection{Study of the MSSM Higgs sector at LC}
The study of the Higgs sector could also provide important information 
concerning possible new physics scenario.
In particular, the MSSM has several characteristic features in the
Higgs sector. First, the Higgs sector is a special type of a two Higgs 
doublet model. The physical Higgs modes contain two CP-even $(h, H)$,
one CP-odd $(A)$, and one pair of charged Higgs bosons $(H^{\pm})$.
There is a theoretical upper bound on the lighter CP-even Higgs boson
$h$, which is about 130 GeV. Since the form of the Higgs potential
is restricted by the requirement of SUSY, the Higgs sector
is parametrized by two parameters at the tree level, usually taken to be
the CP-odd Higgs boson mass $(m_A)$ and the ratio of the two vacuum 
expectation values $(\tan{\beta})$. Although the radiative
correction to the Higgs potential introduces additional parameters like 
the stop mass, Higgs boson masses and couplings are very restrictive, 
and these features are useful in discriminating the MSSM from
other models. 

If only one of CP-even Higgs bosons is found at an earlier stage of 
the $e^+e^-$ LC experiment and the LHC experiment, measurements
of the coupling constants related to the Higgs boson can provide useful
information on model parameters. For instance, the ratios of
the branching ratios such as 
$B(h \rightarrow WW)/B(h \rightarrow \tau\tau)$,
$B(h \rightarrow c \bar{c})/B(h \rightarrow \tau\tau)$
can is deviated from the SM prediction, if the heavy Higgs boson
is not too heavy~\cite{Abe:2001gc,Kamoshita:1995iu, Nakamura:1996wx,
Borisov:1999mu,Carena:2001bg}. 
Indirect information on the heavy Higgs boson mass
may be obtained for $m_A \lsim 600$ GeV, as shown in Figure 4.
On the other hand,
$B(h \rightarrow b\bar{b})/B(h \rightarrow \tau\tau)$ 
should be the same at the tree level in the
MSSM as the SM prediction, but 
the deviation can be large due to SUSY loop corrections to the bottom 
Yukawa coupling constant, especially for a large $\tan{\beta}$
~\cite{Carena:2001bg,Carena:1998gk}.  
\begin{figure}[h]
\raisebox{5cm}{(a)}
\epsfxsize=5cm
\epsfysize=4cm
\epsffile{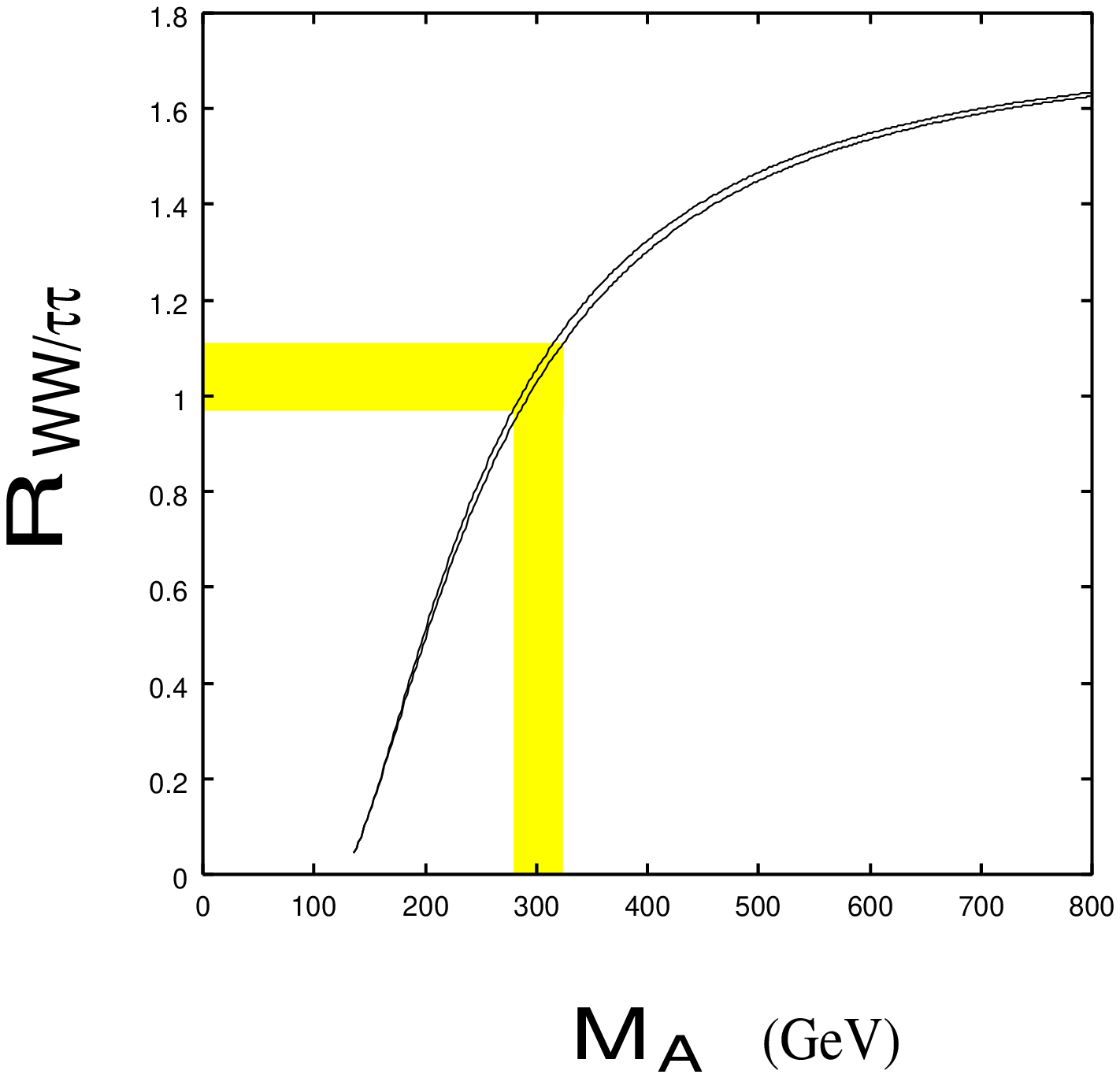}
~~~~~
\raisebox{5cm}{(b)}
\epsfxsize=5cm
\epsfysize=4cm
\epsffile{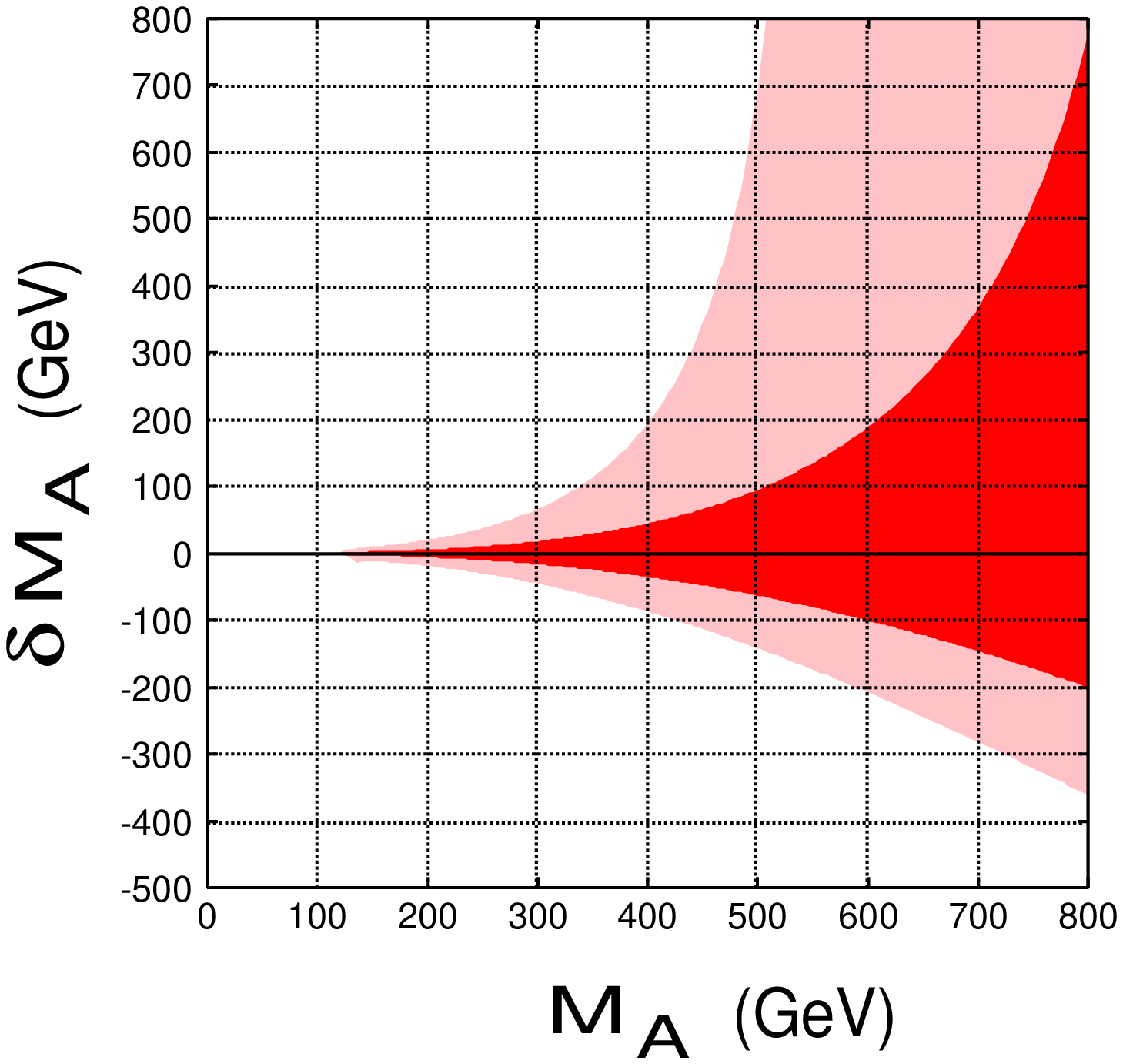} 
\caption{
(a) $B(h \rightarrow W^{(*)}W^{(*)})/B(h \rightarrow \tau\tau)$ vs $m_A$
for the 120 GeV lightest CP-even Higgs boson in the MSSM.
(b)  $m_A$ determination from branching ratios at LC (dark) and
LHC (light) for a 120 GeV Higgs boson. LC corresponds to 
integrated luminosity of 500 fb$^{-1}$. 
$B(h \rightarrow W^{(*)}W^{(*)})/B(h \rightarrow \tau\tau)$
is assumed to be determined with a 15\% error for LHC~\cite{Abe:2001gc}. 
 }
\end{figure}

If the center-of mass energy is large enough, heavy Higgs bosons can be
produced directly. Contours of constant production cross sections 
(0.1 fb$^{-1}$) for various processes in the case of $\sqrt{s}=$ 500 
and 1500 GeV are shown in the parameter space of $m_A$ and 
$\tan{\beta}$ for the MSSM in Figure 5~\cite{Kiyoura:2003tg}. The discovery
limit is essentially determined by the $H$-$A$ and $H^+$-$H^-$ pair production 
threshold except for limited regions of large and small $\tan{\beta}$.
However, heavy Higgs bosons can be seen in different modes 
depending on parameter regions. This will be useful to test the MSSM and
to determine parameters of the model, especially  $\tan{\beta}$.
\begin{figure}[h]
\raisebox{5cm}{(a)}
\epsfxsize=5cm
\epsfysize=4cm
\epsffile{ 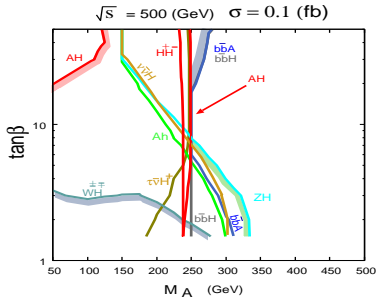}
~~~~~
\raisebox{5cm}{(b)}
\epsfxsize=5cm
\epsfysize=4cm
\epsffile{ 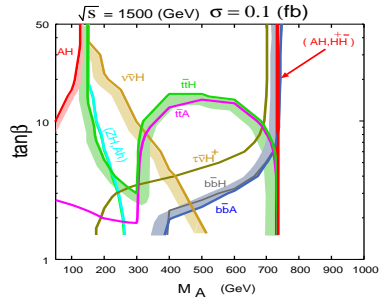} 
\caption{
Contours of production cross sections of 0.1 fb$^{-1}$ for
various heavy Higgs boson processes at LCs with $\sqrt{s}=$ 500 (a) and 
1500 GeV (b)~\cite{Kiyoura:2003tg}. 
 }
\end{figure} 
    
The photon-photon collider option of the LC can be 
especially important for the heavy Higgs searches. This is because
that the s-channel single Higgs production is possible
through the photon-photon-Higgs boson coupling induced at
one-loop level. Since the energy of the back-scattered photon
can be 80 \% of the original electron beam, the heavy Higgs-boson
mass reach can increase up to about 400 GeV for the 500 GeV 
collider. The discovery potential of the heavy Higgs boson
at a photon-photon collider was investigated for$\sqrt{s_{ee}}=$ 630
GeV, and it was shown that the photon-photon collider can cover
the ``wedge'' of the MSSM parameter up to $m_A =500$ GeV,
where only one light Higgs boson is found at 
LHC~\cite{Asner:2001ia}.
The photon-photon collider is also useful to distinguish the CP
property of the Higgs boson using transverse polarization and
angular correlation of decay products $(H \rightarrow t \bar{t}$, for
instance)~\cite{Gunion:1994wy}. The mass reach is, however, 
reduced, if we utilize the transverse polarization. 

\section{Direct Search for New Physics}
Although the SM is very successful so far, there are strong
motivations to search for physics beyond it. First, the SM only
deals with three of the four fundamental interactions, and the gravity is 
not included. Unification of all gauge interactions as well as gauge 
interactions with gravity can be addressed only in the context of physics 
beyond the SM.
Second, the discovery of the neutrino oscillation indicates existence of new
particles and/or interactions. Third, some of fundamental problems 
in cosmology such as what is the dark matter, and how the unbalance 
between baryon and anti-baryon arose require understanding of the physics 
scenario beyond the electroweak scale.

There are many theoretical proposals for physics beyond the SM.
We can expect some signals at the TeV scale, if introduction of new physics
is motivated to explain the separation between the weak scale and the Planck
scale. SUSY and a scenario with large extra-dimensions are 
typical examples. In both case, we will be led to a conceptual change on 
space-time, if it is true.

\subsection{Supersymmetry}
Since the three gauge coupling constants determined at LEP and SLC 
experiments 
turned out to be consistent with the prediction of supersymmetric grand 
unified theory, unified models based on SUSY has been a prime 
candidate of the physic beyond the SM. In SUSY 
models, a superpartner is introduced for each of an ordinary particle. 
New particles include scalar partners of quarks and leptons (squarks 
and sleptons), fermionic superpartners (gluino, charginos, and neutralinos). 
The mass spectrum of these particles depends on SUSY breaking 
scenarios. Therefore, the experimental determination of the spectrum will 
point to a specific scenario. For SUSY search, LHC experiments will provide
a crucial test. In a typical mass spectrum such as one in the minimal 
supergravity model, colored SUSY particles up to 2 TeV can be discovered. 
In addition, a light Higgs boson should exist below about 130 GeV for 
the MSSM, which is an important check point for this model.

Roles of the LC experiment are not restricted to 
discovery of new SUSY particles. Masses, 
quantum numbers, and various couplings of SUSY particles
will be measured in a good accuracy. 
Determination of these quantities without relying on 
some specific model of SUSY breaking is very important to test and 
establish a new symmetry principle of Nature. In many cases of SUSY studies, 
beam polarization can be a very useful tool. This is because 
SUSY is a symmetry between boson and fermion, so that spin information is 
important. Furthermore, the concept of chirality is extended  to the 
squark and slepton sectors in SUSY, so that the initial electron 
polarization is very useful to distinguish left-handed or right-handed 
superparticles.

There are many issues to be addressed on SUSY at LCs.
\begin{itemize}
\item Determination of mass and spin of SUSY particles from decay energy 
distributions, production angle distributions, and threshold scans in pair 
production processes.   
\item Reconstruction of chargino and neutralino mass matrices from 
production cross sections and 
angular distributions with possible effects of CP violation
~\cite{physsusy:Ref:choi.etal}.
\item Determination of the selectron-electron-bino coupling through 
$e^+e^- \rightarrow \tilde{e_R}^+ \tilde{e_R}^-$ for a test of a 
SUSY relation (Figure 6 (a))~\cite{physsusy:Ref:nojiri.etal}.
\item Search for lepton flavor violation in slepton pair production
~\cite{Krasnikov:1995qq, Arkani-Hamed:1996au}.
\item Test of the gaugino mass GUT 
relation (Figure 6 (b))~\cite{physsusy:Ref:tsukamoto.etal}.   
\end{itemize}
These measurements play important roles in determination of SUSY Lagrangian 
in a model-independent way, and proving SUSY. 
\begin{figure}[h]
\raisebox{5cm}{(a)}
\epsfxsize=5cm
\epsfysize=4cm
\epsffile{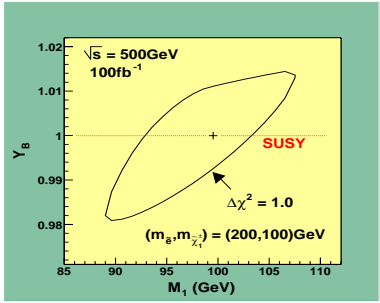} 
~~~
\raisebox{5cm}{(b)}
\epsfxsize=5cm
\epsfysize=4cm
\epsffile{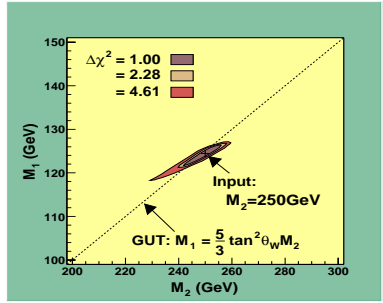}
\caption{Examples of SUSY studies at LCs:
(a) Test of a SUSY
relation from $e^+e^- \rightarrow \tilde{e_R}^+ \tilde{e_R}^-$
~\cite{physsusy:Ref:nojiri.etal}.
(b) Test of the SU(2) and U(1) GUT relation from right-handed selectron
and chargino production processes~\cite{physsusy:Ref:tsukamoto.etal}.
 }
\end{figure} 

In order to obtain a whole picture of a SUSY model, it is most likely that 
information from LHC and LCs has to be combined. For instance, 
the neutralino mass determined at a LC will provide an important input 
for the LHC analysis. Combining colored SUSY particle masses from LHC and 
slepton/chargino/neutralino masses from a LC, we may be able 
to figure out the origin of SUSY 
breaking~\cite{physsusy:Ref:blair.etal}. 

\subsection{Large extra-dimensions}
Recently, a scenario with a large extra-dimensions is proposed, 
motivated by string theory~\cite{Arkani-Hamed:1998rs}.
If there are extra-space dimensions where
only gravity can propagate, the fundamental scale of gravity where all 
interactions are unified can be close to the weak scale. The smallness 
of the gravity can be attributed to the existence of the large extra-space, 
and the Planck scale of $10^{19}$ GeV is not a real physical scale. In this 
case there is no fundamental problem of hierarchy between the Planck and 
weak scales.   

An interesting aspect of this scenario is that we can test 
this idea from collider experiments. Fairly model-independent signals 
of this scenario are emissions of Kaluza-Klein graviton
modes ($e^+e^- \rightarrow \gamma G_{KK}$) and virtual graviton exchange 
in $e^+e^-\rightarrow f \bar{f}$ processes. The sensitivity of LC 
experiments can be comparable to the LHC search for extra-dimensions. 
The cross section of the graviton emission process depends on the 
fundamental scale and the number of extra-dimensions. For a larger number 
of extra-dimensions, the cross section rises more quickly. Experiments 
with different collider energies are useful to determine the number of 
dimensions. Figure 7 illustrates how we can determine the number of 
the extra-dimensions from experiments at LCs with
two different center-of-mass energies~\cite{wilson}. If a signal 
is seen at one beam energy, the energy upgrade is useful to figure out 
the structure of extra-space. By determination of the fundamental scale 
and the number of extra-dimension, we can also derive the size of extra-space. 
The information of the number of the extra-dimensions can be also obtained 
by the analysis of the differential cross section with a fixed center-of-mass 
energy. 
\begin{figure}[ht]
\begin{center}
\leavevmode
\epsfxsize=6cm
\epsfysize=6cm
\epsfbox{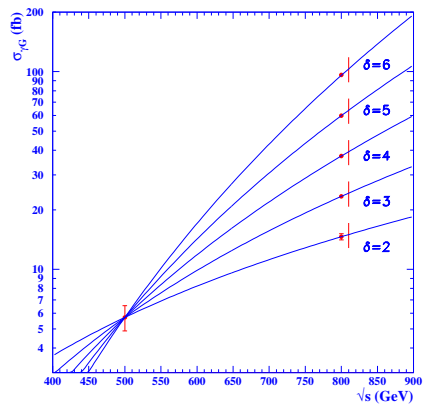}
\caption{
Determination of the number of extra-dimensions at LCs
with the center-of-mass energy of 500 and 800 GeV~\cite{wilson}.
 }
\end{center}
\end{figure}

If some signals of the extra-space are  available within the energy of 
the LC, thresholds of signals of strong gravity or string may 
be open at LHC. These signals include production of black holes
and string resonance states. In order to establish a new picture of 
space-time, both LHC and LCs will be necessary.

\section{Precision studies of top and gauge boson processes}
The studies on the top quark and the W and Z bosons are guaranteed at the 
LC experiment. These studies provides fundamental parameters 
of the particle physics. In addition, we may be able to look for new 
physics effects from precise determination of anomalous coupling constants
involving these particles. 

The top quark is the heaviest particle observed so far. Its production and 
decay are only studies at TEVATRON. An $e^+e^-$ collider can provide a 
unique opportunity to scan the threshold region. We can determine the top 
quark mass at the level of accuracy of 100 MeV and less, and the width
to a few \%~\cite{fms, Martinez-Miquel-2002} . The mass determinaton is 
improved by an order of magnitudes or 
more compared to LHC. The width measurement will be practically only 
possible at the threshold scan of an $e^+e^-$ collider.
An example of the studies at the top production threshold is shown 
in Figure 8.
\begin{figure}[ht]
\begin{center}
\leavevmode
\epsfxsize=10cm
\epsfysize=5cm
\epsfbox{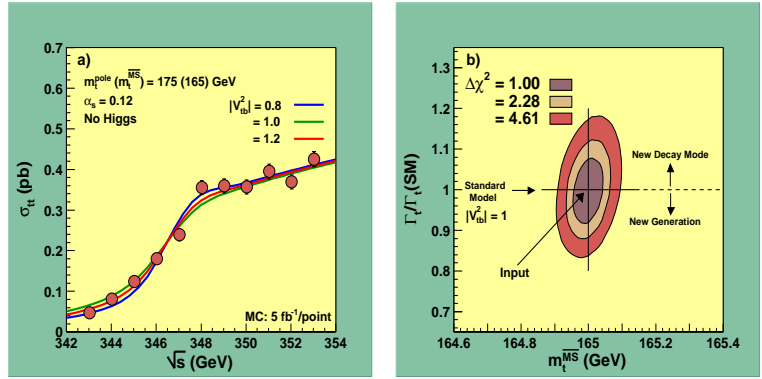}
\caption{Top quark threshold scan for determination of the top mass and 
width \cite{roadmap}.
 }
\end{center}
\end{figure}

The anomalous coupling measurements can be carried out at the threshold 
region as well as at the open-top-production region~\cite{F96}. 
Anomalous couplings 
of the top and photon/Z boson include the top electric dipole 
moment which is induced by CP violation outside the SM.
The accuracies of electronic dipole moments determinations are 
$10^{-19}$-$10^{-18}$ e cm, for instance. At LHC, information on 
the gluon anomalous coupling will be available.

Triple and quadratic anomalous couplings of gauge bosons will be also 
measured precisely at LCs. Although constraints on these coupling 
constants are already obtained by LEP and TEVATRON experiments,
the higher energy and beam polarization are advantages of the LC 
experiment for this study. These measurements are particularly important 
when a light Higgs boson is not found at the LHC and LC 
experiments, in which case something is missing in our current understanding
of the electroweak symmetry breaking. 
Sensitivities to triple gauge boson coupling constants $\kappa_V$
and $\lambda_V$
defined in
$
{{\cal L}_{WWV}}/g_{WWV} 
 =     i g^V_1 ( W_{\mu\nu}^\dagger W^\mu V^\nu 
            - W_\mu^\dagger V_\nu W^{\mu\nu} ) 
 + i\kappa_V W^\dagger_\mu W_\nu V^{\mu\nu} 
 + { i\lambda_V \over m^2_W } 
     W^\dagger_{\lambda\mu} W^\mu_\nu V^{\nu\lambda} 
$
are $10^{-4}$-$10^{-3}$ at LC experiments
~\cite{Abe:2001wn, Aguilar-Saavedra:2001rg, Abe:2001gc}.

\section{Summary}
Current understanding of elementary particle physics is based on the 
SM, in which the gauge principle and the electroweak symmetry 
breaking are two basic ingredients. Nature of gauge interactions has 
been studied experimentally in full details last twenty years, but we know
little about the electroweak symmetry breaking and the Higgs mechanism. 
The most important next step of the high energy physics is to clarify the 
dynamics of electroweak symmetry breaking and the mass generation mechanism 
of elementary particles. For this purpose, an $e^+e^-$ LC
is necessary. The LC with the center-of-mass energy of up to 
500 GeV and an integrated luminosity of 500 fb$^{-1}$ can be considered as 
a Higgs factory. We can determine the coupling constants related to the mass 
generation mechanism very precisely. In order to determine the top
Yukawa coupling constant and the Higgs self-coupling constant with good 
precisions, a higher center-of-mass energy is required. 
The LC experiments will also play essential roles in finding 
and studying new physics signals, such as SUSY and the scenario 
with a large extra-dimensional space. For these analysis, the LHC and linear 
collider experiments can be complementary to each other. The clean 
experimental 
environment, availability of the beam polarization, capability of energy scan
at the LC is very useful for these studies. In addition to 
direct search for new signals, precise studies on the SM 
processes including the top quark and the W boson provide  alternative 
ways to look for new physics effects.

The high energy physics has been developed by concurrent running 
of hadron and lepton machines. Since the end of the LEP operation, there is 
no energy frontier $e^+e^-$ machine. There are very strong physics cases 
to construct the $e^+e^-$ LC which can operate concurrently 
with the LHC experiment. The physics of the TeV energy scale may or may not 
be what we are expecting, but whatever it may be, it will be necessary for 
us to have both machines to elucidate the physics at the electroweak 
scale and beyond.

\end{document}